\documentclass[namedreferences]{SolarPhysics}
\usepackage{graphicx}        
\usepackage{amssymb}        
\usepackage{url}             
\newcommand{\beq}{\begin{equation}}
\newcommand{\eeq}{\end{equation}}
\newcommand{\eq}[1]{(\ref{#1})}
\newcommand{\fb}{}

\newcommand{\aap}{    {\it Astron. Astrophys.}}

\newcommand{\mnras}{  {\it Mon. Not. Roy. Astron. Soc.}}

\newcommand{\solphys}{{\it Solar Phys.}}

\def\copyrightline{}

\begin{document}
\begin{article}
\begin{opening}
\title{Width of Sunspot Generating Zone\\
and Reconstruction of Butterfly Diagram}
\author{V.~G.~\surname{Ivanov}$^{1,2}$}
\author{E.~V.~\surname{Miletsky}$^1$}
\runningauthor{Ivanov and Miletsky}
\runningtitle{Width of Sunspot Generating Zone and Reconstruction of Butterfly Diagram}
\institute{$^{1}$ Central Astronomical Observatory at Pulkovo, Saint-Petersburg, Russia\\
$^2$ email: \url{ivanov.vg@gao.spb.ru}}
\begin{abstract}
Based on the extended Greenwich--NOAA/USAF catalogue of sunspot
groups it is demonstrated that the parameters describing the
latitudinal width of the sunspot generating zone (SGZ) are closely
related to the current level of solar activity, and the growth of
the activity leads to the expansion of SGZ.  The ratio of the
sunspot number to the width of SGZ shows saturation at a
certain level of the sunspot number, and above this level  the
increase of the activity takes place mostly due to the expansion of
SGZ. It is shown that the mean latitudes of sunspots can be
reconstructed from the amplitudes of solar activity. Using the
obtained relations and the group sunspot numbers by \inlinecite{hs},
the latitude distribution of sunspot groups (``the
Maunder butterfly diagram'') for the 18th and the first half of the
19th centuries is reconstructed and compared with historical sunspot
observations.
\end{abstract}
\keywords{Solar cycle, observations; Sunspots, statistics}
\end{opening}

\section{Introduction}

The 11-year cycle of solar activity, which is now referred to as
``the Schwabe-Wolf law'' after the names of its discoverers
(Schwabe, 1843, 1844; Wolf, 1852), was found in the study of sunspot
number variations in time. Later a gradual shift of the mean
latitude of the sunspot generating zone (SGZ) to the solar equator
in the course of the 11-year cycle was found \cite{ref4,ref5}. This
spatial regularity of the cycle is now called ``the Sp\"orer law''.
The famous ``butterfly diagram'', first drawn by Maunder
(\citeyear{mau}), can be regarded as a graphic illustration of this
law.

Since sunspots are among the most prominent manifestations of the
solar magnetic field, these two laws, apparently, reflect different
aspects of the solar magnetic cycle. Therefore, for better
understanding of the physical nature of the 11-year cycle it is very
important to explore the relations between the latitudinal
distribution of sunspots and solar activity.

A close connection is known between the sunspot activity indices (in
particular, the Wolf number W) at the maximum of a solar cycle with
such characteristics as the sunspot latitudes averaged over the year
of the solar maximum \cite{ref7} or over the solar cycle
\cite{ref8}. Recently \citeauthor{ar09} (\citeyear{ar09}, hereafter
referred to as Paper~I) confirmed this connection by reproducing the
Waldmeier relation with new data on sunspot activity obtained after
1955. In investigation of statistical characteristics of the
latitudinal distribution of sunspots in different cycles it was also
found that the mean solar latitude of a given cycle and their
latitudinal dispersion are highly correlated \cite{ref9}. Therefore,
latitudinal characteristics of the 11-year cycles taken as a whole
are related closely to the power of these cycles.

It is interesting to look for the relation between the latitudinal
sunspot distribution and the instantaneous level of solar activity.
One of such possible candidates  is the latitudinal width of SGZ. It
was found in the 1950s by \inlinecite{ref10} and \inlinecite{ref11}
that a quantity which they introduced as a measure of the SGZ width
depends upon the phase of the 11-year cycle and its largest values
fall on the epoch of the maximum of the cycle. However, they did not
find any dependence of this quantity upon either the magnitude of
the 11-year cycle or the current level of solar activity.

In Paper~I we found that the yearly means of the SGZ width
(SZW), expressed as the difference of maximal and minimal latitudes
of sunspot groups, are tightly related to the level of sunspot
activity characterized by the Wolf numbers. In the present paper we
continue this study, considering other characteristics of SZW and
sunspot activity indices and using both yearly and solar rotation
means of the corresponding values.
We will also demonstrate that behavior of the activity can be used to
find the mean latitude of the SGZ.
Using the obtained regularities
we will develop an approach which allows us to reconstruct the
latitudinal parameters of sunspot distribution (and, thereby, to
restore the butterfly diagram) before the middle of the 19th century
on the basis of the available information about the amplitudes of
the activity.

\section{Indices of the Number and  Latitudinal Distribution of Sunspots}
\label{dsigma}
\subsection{Data and Indices}
\label{sec.ind}
Using the data on sunspot latitudes from the Greenwich catalogue and its NOAA/ USAF extension%
\footnote{\url{http://solarscience.msfc.nasa.gov/greenwch.shtml}}
for the epoch 1874--2006, we calculate daily values of several
sunspot indices. In particular, we obtain the daily numbers of
sunspot groups (G), total sunspot areas (SA), and the mean latitudes
of sunspots weighted with the sunspot area (LA). For the same data
we find the standard deviations of sunspot group latitudes
($\sigma$) and the highest (LAH) and lowest (LAL) latitudes of
sunspots for a given day. When we calculate the latter quantities,
we consider the highest latitude to be equal to the lowest if there
is only one sunspot group in a given hemisphere.

In Paper~I we used indices LAH and LAL as characteristics of the
upper and lower boundaries of SGZ, and the derived index ${\rm
D}={\rm LAL}-{\rm LAH}$ (the difference between the highest and
lowest latitudes, {\it i.e.} the latitudinal size of the wings of
the butterfly diagram) as a measure of SZW. In the present paper we
use also another possible measures of SZW: the daily standard
deviations of sunspot latitudes $\sigma$ and the extent of the zone
$\Delta\phi_\rho$ on a given level of the sunspot density $\rho$
(which will be discussed in Sec.~\ref{bpars}).

In the following we use yearly and solar rotation means of indices,
so that the above mentioned daily values will be averaged over the
corresponding time ranges.

\begin{figure}
\begin{center}
\fb{\includegraphics[width=\textwidth]{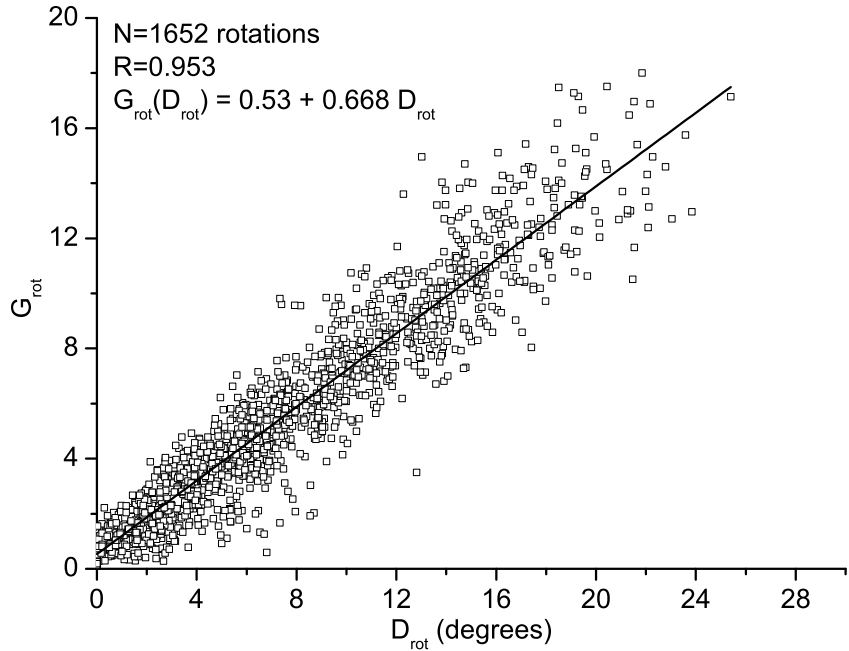}}
\end{center}
\caption{%
The relation between the rotation means of the sunspot group index ${\rm G}_{\rm rot}$ and
the measure of the SGZ width ${\rm D}_{\rm rot}$.}
\label{fig1}
\end{figure}

\subsection{Relations between the Amplitude and Latitudinal Indices of Sunspots}

In the interval 1874--2006 (133 years) the yearly indices of sunspot
activity (the Wolf number W, sunspot group index G, and their total
area SA) are proved to be strongly correlated to all the aforenamed
measures of SZW both for global and hemispheric values of these
indices (see Table~\ref{tab1}). In particular, as it was shown in
Paper~I, the correlation coefficient $R({\rm W}, {\rm D})=0.975$. It
is remarkable that the maximum value in Table~\ref{tab1} $R({\rm G},
{\rm D})=0.989$ corresponds to the correlation between an activity
amplitude index and a latitudinal index, whereas the highest
correlation within the group of activity amplitude indices (G, W,
SA), which are commonly believed to be highly correlated, is lower
($R({\rm W}, {\rm SA})=0.985$). One should note that for 133 points
for which the correlations were calculated, their confidence
probabilities (for $R=0.989$) are over three standard deviations. We
can see also that the relations between alternative measures of SZW
are high enough, the highest being between $D$ and $\Delta\phi^2$.

\begin{table}
\begin{center}
\begin{tabular}{cccccccccc}
\hline
                      &     G&     W&     SA&      D&  D$^2$& $\sigma$& $\sigma^2$& $\Delta\phi_{0.03}$& $\Delta\phi^2_{0.03}$\\
\hline
G&                    & 0.982& 0.971&  \textbf{0.989}&  0.944&  0.920&  0.933&  0.936&  0.970\\
W&                    &      & 0.985&  0.975&  0.944&  0.912&  0.936&  0.914&  0.955\\
SA&                   &      &      &  0.958&  0.926&  0.889&  0.913&  0.898&  0.938\\
D&                    &      &      &       &  0.956&  0.949&  0.965&  0.943&  0.980\\
D$^2$&                &      &      &       &       &  0.858&  0.936&  0.845&  0.928\\
$\sigma$&             &      &      &       &       &       &  0.969&  0.944&  0.946\\
$\sigma^2$&           &      &      &       &       &       &       &  0.904&  0.947\\
$\Delta\phi_{0.03}$&  &      &      &       &       &       &       &       &  0.979\\
\hline
\end{tabular}
\caption{%
Correlations between the yearly means of sunspot activity indices and various measures of SZW.}
\label{tab1}
\end{center}
\end{table}

To examine how the relation between sunspot activity and SZW depends
on the range of averaging, we calculated the correlation between
solar rotation means ${\rm G}_{\rm rot}$ and ${\rm D}_{\rm rot}$ of
the indices G and D for the same epoch 1874--2006 (1652~solar
rotations). One can see in Figure~\ref{fig1} that the obtained
dependence can be fairly well presented by the linear relation
\begin{displaymath}
{\rm G}_{\rm rot} = 0.53+0.668\,{\rm D}_{\rm rot} \quad (R=0.953)
\end{displaymath}
and has high consistency which follows from uniform distribution of
points over the ranges of the indices. A decrease in the correlation
as compared with the value obtained for yearly means of the indices
is as small as 0.036 and the mean square error between the observed
and model series is $\delta({\rm G}_{\rm rot})=1.18$.

\begin{figure}
\begin{center}
\fb{\includegraphics[width=\textwidth]{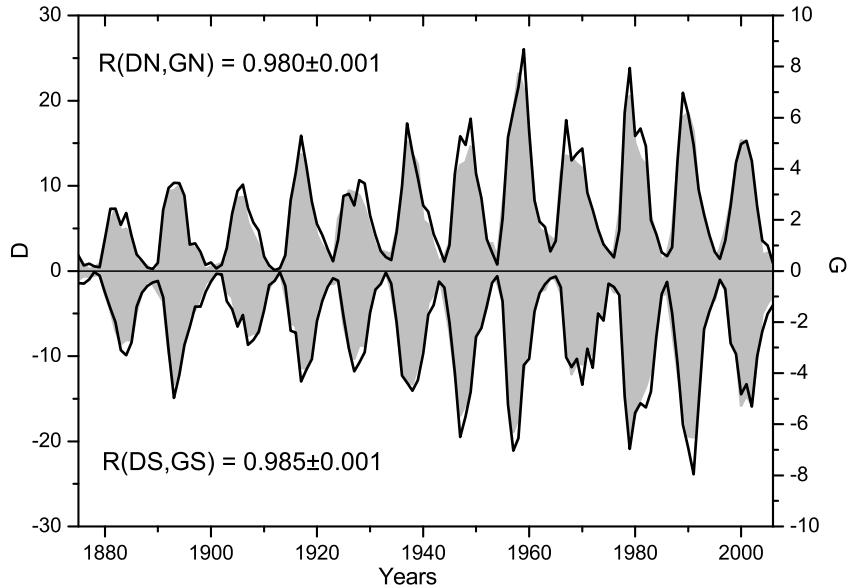}}
\end{center}
\caption{%
Hemisphere components of indices G and D. The series ${\rm GN}$ (solid line)
and ${\rm DN}$ (dotted line), corresponding to the northern hemisphere,
are plotted to the positive direction of the ordinate, and ${\rm GS}$ and ${\rm DS}$
(corresponding to the southern hemisphere) to the negative direction.}
\label{fig2}
\end{figure}

The behavior of sunspot indices and SZW taken separately in
different hemispheres are also proved to be in good agreement. For
the yearly means of indices G and D in the northern (N) and southern
(S) hemispheres one can obtain the following relations:
\begin{displaymath}
{\rm GN}({\rm DN}) = 0.052+0.355\,{\rm DN} \quad (R({\rm DN},
{\rm GN})=0.980, \quad
\delta({\rm GN}) = 0.39)
\end{displaymath}
and
\begin{displaymath}
{\rm GS}({\rm DS}) = -0.001+0.360\,{\rm DS} \quad
(R({\rm DS},{\rm GS})=0.985, \quad
\delta({\rm GS}) = 0.33).
\end{displaymath}
In Figure~\ref{fig2} the hemispheric indices are presented. The
series ${\rm GN}$ and ${\rm DN}$ are shown by the solid and dotted
lines, respectively, and plotted to the positive direction of the
ordinate. Similarly, ${\rm GS}$ and ${\rm DS}$ are plotted to the
negative direction of the ordinate. Some disagreements between the
behavior of indices G and D take place only during the minima of
11-year cycles, but in other epochs of the cycles the indices agree
very well.

Recently (Paper~I) we showed that the yearly means of indices D and
W are closely related. The above examination proves that this
regularity holds for other characteristics of SZW and activity
indices, and for different ranges of averaging. Therefore, the
growth of solar activity is accompanied by an increase of SZW. Using
this relation and having the latitudes of sunspots (and SZW), one is
able to make a reliable estimation of the level of solar activity.
Conversely, by knowing this level, one can calculate rather
accurately the current extent of ``the wings of Maunder's
butterfly'' (see Section~\ref{brec}).

\begin{figure}
\begin{center}
\fb{\includegraphics[bb=20 20 265 202]{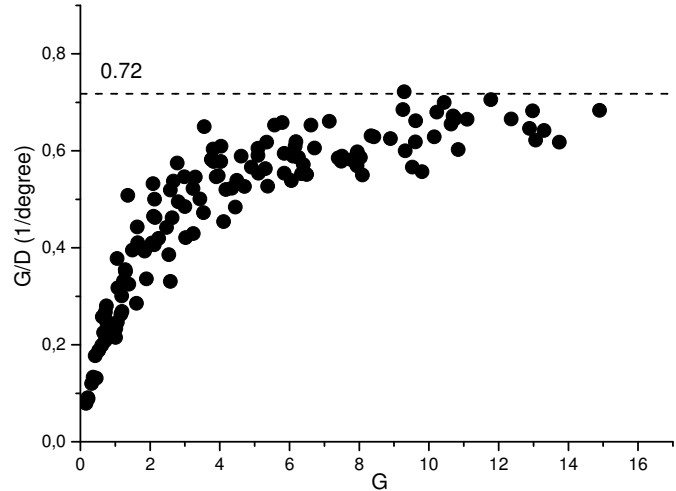}}
\end{center}
\caption{%
The dependence of the G/D ratio  upon G (yearly means). The dashed line corresponds to the saturation level 0.72.}
\label{fig3}
\end{figure}

\subsection{Saturation Level of G/D}

The ratio G/D of the sunspot group index (G) to the width of the
corresponding zone (D) can be, under certain conditions, treated as
a mean density of the latitudinal distribution of sunspots. In
Figure~\ref{fig3} the dependence of this value upon G is plotted. We
can see that, as G increases, a linear growth of the ratio G/D slows
down, and saturates at a certain level ($\approx 0.72$).

\begin{figure}
\begin{center}
\fb{\includegraphics[bb=20 20 292 205]{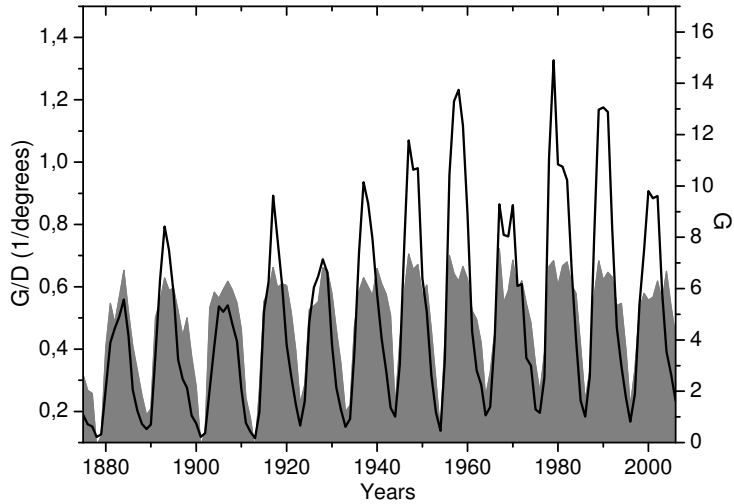}}
\end{center}
\caption{%
The yearly means of the ratio G/D (the dotted line) and the group number index G (the solid line).}
\label{fig4}
\end{figure}

A comparison of the behavior of the yearly means of G/D (see
Figure~\ref{fig4}) shows that, in each 11-year cycle, this value
starts increasing in the minimum and reaches a level which is nearly
the same for all 11-year cycles. A further increase of the activity
takes place mostly due to the widening of SGZ in latitude.
Therefore, there is a limitation of the mean latitudinal density of
sunspots.

\section{Reconstruction of the Butterfly Diagram}
\label{brec}

Systematic data on the coordinates of sunspot groups are available
in the extended Greenwich catalogue since 1874. Earlier information
on the spatial distribution of sunspots is scarce, however. For
1854--1873 the mean latitudes of sunspots can be obtained by the
compilation of pre-Greenwich observations
\cite{esai}%
\footnote{See ESAI database:
\url{http://www.gao.spb.ru/database/esai/}}.
But before the middle
of the 19th century we have only non-systematic drawing of the Sun
({\it e.g.}, \citeauthor{ribes}, 1993; \citeauthor{arlt08}, 2008,
2009).

The relation between the amplitude of solar activity and SZW found
above can help us to reconstruct the latitudinal distribution of
sunspots in the pre-Greenwich epoch from some proxy data. We will
use as such proxy the group sunspot numbers (GSN), which were
calculated by \citeauthor{hs} (1998) since 1610. Its correlation
with the sunspot group index G on the overlapping period of data is
higher than 0.99 for yearly means, and the linear relation
\begin{equation}\label{g118}
{\rm G} \approx {\rm GSN} / 11.76
\end{equation}
is valid for these two indices. Therefore, in the following we will
not make a difference between GSN (rescaled according to
Equation~\eq{g118}) and G.

\begin{figure}
\begin{center}
\fb{\includegraphics[width=0.6\textwidth,bb=-30 180 625 680]{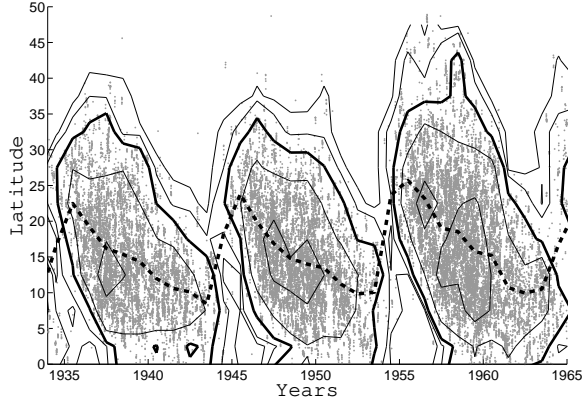}}
\end{center}
\caption{%
A portion of the butterfly diagram. The levels of the sunspot group densities are
shown by solid contours, with the bold contours corresponding to the
density level $\rho=0.03\;$groups/degree/day. The dashed line shows
the mean latitude of sunspot groups.} \label{fig.params}
\end{figure}

\subsection{Parametrization of the Butterfly Diagram}
\label{bpars}

In order to reconstruct the spatial distribution of sunspots from a
scalar index, one first should select a proper parametrization of
the distribution. The latitudinal distribution of sunspot groups in
the butterfly diagram for a given year in a given hemisphere can be
approximately described, {\it e.g.}, by the mean latitude of
sunspots $\phi_0$ and some measure of SZW. As the latter, it is
possible to use any of the SZW indices mentioned in
Section~\ref{sec.ind}, but for our purpose it is convenient to use
the extents of the zone $\Delta\phi_\rho$.

To determine it we plot the contours of latitudinal sunspot group
densities on the  butterfly diagram, and select a ``representative''
contour which outlines the wing of the butterflies fairly well (see
Figure~\ref{fig.params}). In the following we select
$\rho=0.03\;$groups/degrees/day (the thick line in
Figure~\ref{fig.params}) and will omit the index $\rho$ on
$\Delta\phi_\rho$. The upper $\phi_{\rm up}$ and lower $\phi_{\rm
low}$ borders of the wing in a given hemispheres are determined as
the yearly averages of latitudes of the corresponding contours, and
the half-widths of the wing in the hemisphere, as $\Delta\phi_{\rm
up} = \phi_{\rm up}-\phi_0$ and $\Delta\phi_{\rm low} = \phi_0 -
\phi_{\rm low}$.

Common sunspot activity indices, such as GSN, which we use for the
reconstruction of the sunspot distribution, do not include explicit
information of the north-south asymmetry of sunspots. Therefore, we
will neglect the asymmetry, coming to the values of these parameters
averaged over two hemispheres.

Of course, in case of need, by selecting an additional level of the
density $\rho'$, one can obtain the corresponding indices
$\phi'_{\rm up, low}$ and parametrize the latitudinal distribution
of sunspots in more detail.

\begin{figure}
\begin{center}
\fb{\includegraphics[bb=22 20 300 220]{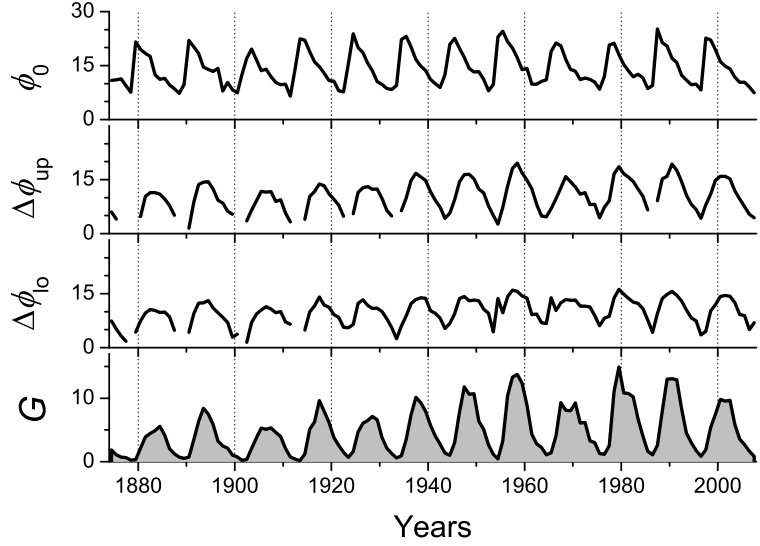}}
\end{center}
\caption{%
From top to bottom: the mean latitudes of sunspot groups $\phi_0$,
the half-widths of the butterfly diagram $\Delta\phi_{\rm up}$ and
$\Delta\phi_{\rm low}$, and the number of sunspot groups G.}
\label{fig.idx}
\end{figure}

Therefore, we have selected three yearly series ($\phi_0$,
$\Delta\phi_{\rm up}$, and $\Delta\phi_{\rm low}$), which describe
the form of the butterfly diagram (see Figure~\ref{fig.idx}), and
will look for their relations to the level of solar activity
described by G.

\begin{figure}
\begin{center}
\fb{\includegraphics[width=0.47\textwidth,bb=20 20 290 222]{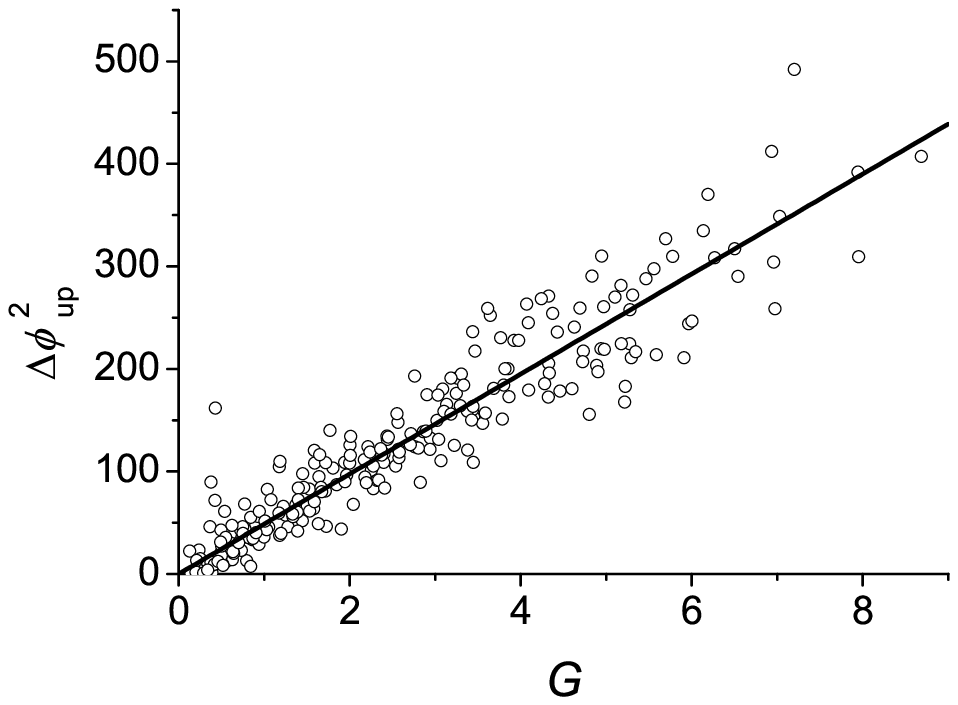}}
\fb{\includegraphics[width=0.47\textwidth,bb=20 20 290 222]{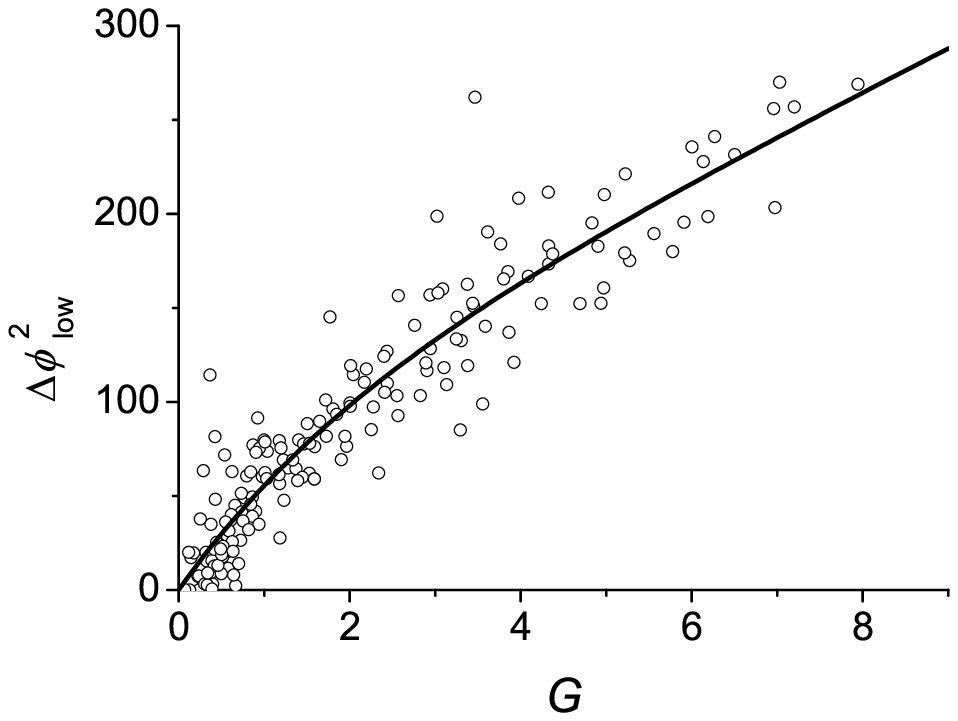}}
\end{center}
\caption{%
The relationship between the number of sunspot groups G and the
half-widths of the butterfly diagram $\Delta\phi_{\rm up}$ and
$\Delta\phi_{\rm low}$.} \label{fig.gdphi}
\end{figure}

\subsection{Reconstruction of the Half-Widths of the Butterfly Diagram}

The average half-width of the butterfly wings
$\Delta\phi=(\Delta\phi_{\rm up}+\Delta\phi_{\rm low})/2$ can be
used as a measure of the latitudinal extent of SGZ. Table~\ref{tab1}
shows that its square $\Delta\phi^2$ is in good correlation with G.
However, to improve the accuracy it is useful to treat
$\Delta\phi_{\rm up}$ and $\Delta\phi_{\rm low}$ separately. In
Figure~\ref{fig.gdphi} one can see that their dependence upon $\sqrt
{\rm G}$ is close to linear and can be described by the regression
equations
\begin{equation}
\Delta\phi_{\rm up}({\rm G}) = \alpha \; \sqrt{{\rm G}}
\end{equation}
and
\begin{equation}
\Delta\phi_{\rm low}({\rm G}) = \beta \; \sqrt{ {\rm G} + \gamma (1 - \exp(-{\rm G}/2))}
\end{equation}
where
\[
\alpha=6.98\,, \qquad
\beta=4.79\,, \qquad
\gamma=3.61\,.
\]
The correlation coefficients for both regressions are 0.94 and the
standard errors are about $1.5^\circ$. We introduced a week
non-linear tuning by parameter $\gamma$ in the second relation to
improve the fitting at low G.

\subsection{Reconstruction of the Mean Latitudes of Sunspots}

\begin{figure}
\begin{center}
\fb{\includegraphics[width=0.5\textwidth,bb=20 20 287 219]{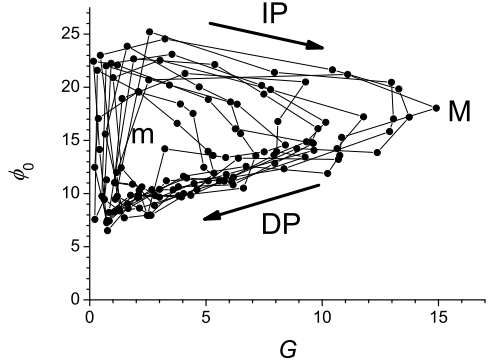}}
\end{center}
\caption{%
The relations between the yearly index G and the mean latitude of sunspots $\phi_0$ at various phases of the 11-year cycle: the minimum (m), increasing phase (IP), maximum (M), and decreasing phase (DP).}
\label{fig.dphi}
\end{figure}

The reconstruction of the mean sunspot latitudes is a more difficult
problem, since there is no univocal relationship between this index
and solar activity indices. Figure~\ref{fig.dphi} shows that the
relation between the mean sunspot latitude and G
differs on different phases of the 11-year solar cycle.
Therefore, in order to find $\phi_0(t)$ for a given year $t$, one
should take into account not only G($t$), but also the time
derivatives of this index (or equivalently, G($t'$) for $t'\ne t$).

Nagovitsyn (\citeyear{nag}) proposed a method for the reconstruction
of the mean latitudes which is based on the mapping of an original
characteristic of solar activity $X$ into a pseudo-phase space
\begin{equation}
\label{dps}
X(t) \longrightarrow \Psi(t) =
[ X(t - n\cdot\Delta),  X(t - (n - 1)\cdot\Delta), \dots , X(t), \dots , X(t+n\cdot\Delta) ]
\end{equation}
and search for a linear regression which links $\Psi$ and the
required index $\phi_0$. Using this method, he obtained a
reconstruction of the mean latitudes of sunspots since 1621.

However, in this approach it is difficult to estimate the errors of
the model. Below we apply another method, which does not assume
linear relationships. For the reconstruction we use an artificial
two-layers feed-forward neural network (see, {\it e.g.},
\citeauthor{conway} (1998) and references therein) with sigmoid
(linear) transfer functions of the first (second) layer,
respectively. The input variable of the neural network (NN) is the
group index G mapped into the pseudo-phase space by Equation
\eq{dps} (with $\Delta=1\;$year) and the output variable is a
20-dimensional vector which is made of the yearly means of sunspot
group densities in the five-degree latitudinal intervals
$[-50^\circ,-45^\circ]$, $\dots$, $[+45^\circ,+50^\circ]$. We are
free to vary three parameters of the model, namely the dimension of
the input vector $2n+1$, the number of neurons of the first hidden
layer $h$, and the initial state of NN.

\begin{figure}
\begin{center}
\fb{\includegraphics[width=\textwidth,bb=20 20 317 121]{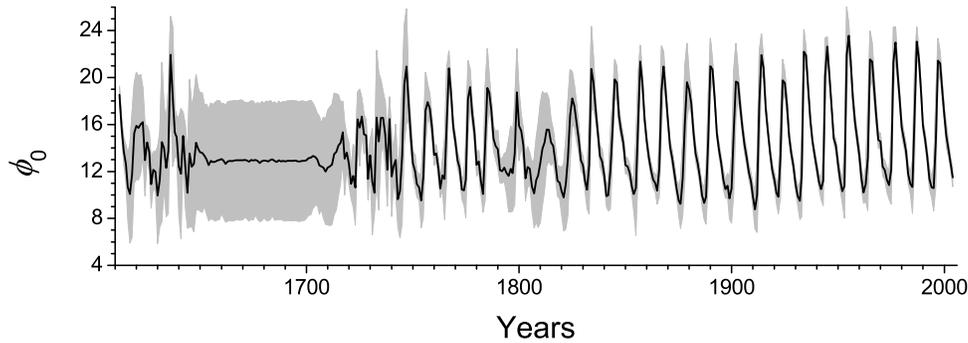}}
\end{center}
\caption{%
The reconstruction of the mean sunspot latitudes $\phi_0$
(1612-2004) by the NN model. The gray halftone corresponds to the
error of reconstruction $\delta\phi_0$.} \label{fig.phi0}
\end{figure}

\begin{figure}
\begin{center}
\fb{\includegraphics[width=0.9\textwidth,bb=20 20 276 113]{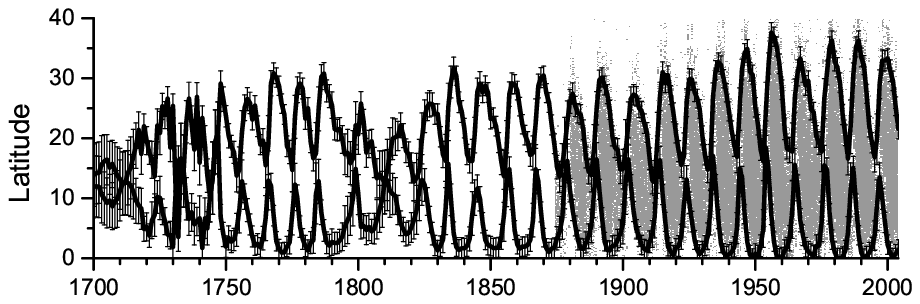}}
\fb{\includegraphics[width=0.9\textwidth,bb=20 20 276 113]{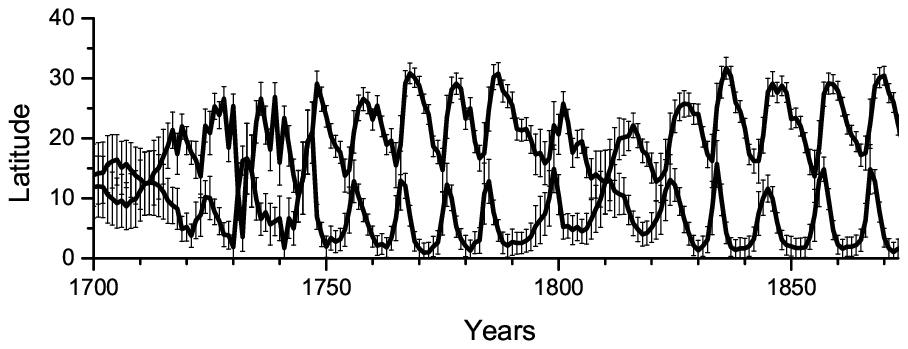}}
\end{center}
\caption{%
The reconstruction of the northern wing of the butterfly diagram. The full
series for 1700--2004 years is shown on the top panel, and the reconstruction
for 1700--1874 on the bottom panel. The vertical bars correspond to the errors
of the reconstruction. The gray halftone on the top panel is the observed sunspot distribution.
}
\label{fig.wing}
\end{figure}

Selecting a random set of these parameters (under the condition $2
\le n \le 23$ and $2 \le h \le 11$) and training each of the
corresponding NN, we obtain an ensemble of $Q$ different models.
Then we use as the input of these models GSN by Hoyt and Schatten
(1998) recalculated to the G scale by Equation \eq{g118} and mapped
into the pseudo-phase space. For each of the resulting output
density distributions we calculate a series of mean sunspot
latitudes $\phi_{0,i}, i=1,\dots,Q$. Finally, the resulting
reconstruction of $\phi_0$ (Figure~\ref{fig.phi0}) is obtained by
averaging these series over the ensemble, and the corresponding
standard deviation $\delta\phi_0$ can be treated as an estimate of
the error in the method. A comparison between the mean latitudes
obtained by \inlinecite{nag} and our reconstruction shows that the
series are in fair agreement. However, the errors in our
reconstruction for low levels of the global solar activity are
significantly larger. It can be explained by the fact that, in the
epoch of the Maunder minimum, the system was located in a domain of
the pseudo-phase space which differs from the one used for the
training of NN. Therefore, the mean latitude can be more or less
reliably reconstructed by this method only after the beginning of
the 18th century.

\subsection{Results of Reconstruction}
The reconstructed parameters $\phi_0$, $\phi_{\rm up}$, and $\phi_{\rm
low}$ can be used to restore the form of the butterfly diagram in
the 18th and the first half of the 19th centuries
(Figure~\ref{fig.wing}).

\begin{figure}
\begin{center}
\fb{\includegraphics[width=0.75\textwidth,bb=47 56 372 267,clip]{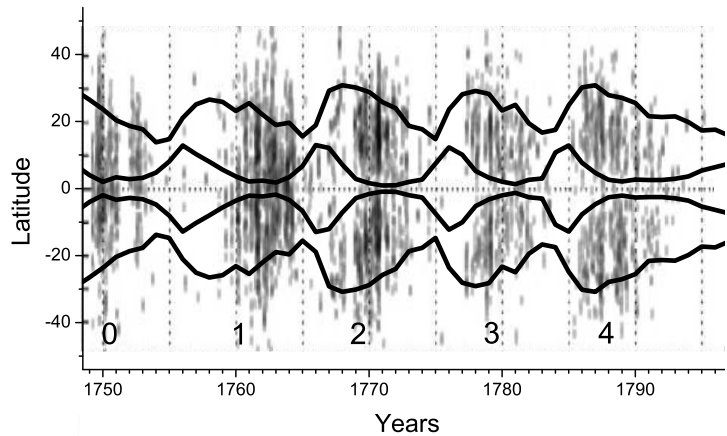}}
\end{center}
\caption{%
A portion of our reconstruction (the bold lines correspond to $\phi_{\rm up}$ and $\phi_{\rm low}$) superposed on the distribution of sunspots by Staudacher \protect\cite[Figure~2a]{arlt09}. The numerals below are  the cycle numbers.}
\label{fig.arlt}
\end{figure}

It is interesting to compare the obtained diagram with observations.
Unfortunately, the largest set of sunspot drawings made in Paris
Observatory \cite{ribes} was collected in the earlier epoch
(1660--1719). Another long series of sunspot coordinates was
recently extracted from semi-centennial (1749--1796) observations of
a German amateur astronomer Staudacher \cite{arlt09}. In
Figure~\ref{fig.arlt} the butterfly diagram by the data of
Staudacher (the upper panel of Figure~2 from \inlinecite{arlt09}, which
contains 6285 sunspot positions) is compared with $\Delta\phi_{\rm
up}$ and $\Delta\phi_{\rm low}$ of our reconstruction. One can see
that for cycles 3 and 4 the distributions are more or less in
agreement. For the earlier three cycles the form and size of the
wings visibly differ from our reconstruction. However, as it is
noted by \inlinecite{arlt09}, the distributions in cycles 0-2 have some
anomalies: the excess of sunspots near the solar equator, unclear
equatorward migration of sunspots during the cycle, etc. It is not
clear whether these features correspond to a real behavior of the
Sun or it is an artifact caused by methods of observation or data
processing. The data of Staudacher are not uniformly distributed
over time (during cycles 0-3 observations are more frequent than
later). Besides, possibly, the data are affected by ``the factor of
attention'', {\it i.e.} more frequent observations in days with
larger numbers of sunspots as compared with days with few or no
sunspots (for 1016 days with observations there are only 17 days
without sunspots). This factor would obviously lead to systematic
overestimation of SZW.

Another possible reason for the peculiar form of cycles 0-3, as it
was also mentioned by Arlt (\citeyear{arlt09}), is the dominance of
the quadrupolar mode of the magnetic field in this epoch. Of course,
this effect cannot be reproduced by our model, which is built under
the assumption of the dipolar butterfly-like form of the sunspot
distribution.

Taking into account possibility of such effects, we consider the
agreement of these two distributions to be satisfactory.

\section{Conclusions}
In this paper we continued the investigation of Paper I on the
relationship between the level of solar activity and some
characteristics of the latitudinal distributions of sunspots. In
particular, we showed that in the 11-year cycle the characteristic
width of the sunspot generating zone is tightly related to the level
of solar activity, and this relation holds for different indices of
sunspot activity, regardless of the selection of parameters to
define the extent of latitudinal sunspot distribution or
various scales of averaging.

We found that a certain saturation level exists for the ratio (G/D)
of the sunspot number index (G) to the latitudinal size of the
corresponding zone (D). Above this level, the increase of the
activity takes place mostly due to the expansion of SGZ. In all the
explored sunspot cycles the ratio reaches this level and, therefore,
practically does not depend upon the amplitudes of the cycles.

We also showed that the mean latitude of the sunspot distribution is
related to the levels of the activity in the given sunspot cycle

Using the obtained relation we reconstructed the form of the
latitudinal distribution of sunspots (``the Maunder butterfly
diagram'') in the epochs where little or no direct observations of
spatial sunspot distribution are available. The reconstruction is
rather accurate for middle or high levels of global solar activity
(1720-1863), but during the epochs of grand minima (in the Maunder
minimum) its results are not reliable.

Our results can be regarded as an additional confirmation of
complementarity of spatial and amplitude characteristics of solar
magnetic fields. The obtained regularities can be used as diagnostic
criteria for the choice of adequate models of the solar cyclicity.

\section{Acknowledgements}

This work is supported in part by the grant of Russian Foundation
for Basic Research No.~10-02-00391 and the grant ``Leading
Scientific Schools'' No.~3645.2010.2. The authors are also grateful
to Yu.~Nagovitsyn for useful discussions.

\end{article}
\end{document}